\documentstyle[12pt]{article}

\def\hybrid{\topmargin 0pt      \oddsidemargin 0pt
        \headheight 0pt \headsep 0pt

       \textwidth 6.5in        % US paper
       \textheight 9in         % US paper
        \marginparwidth 0.0in
        \parskip 5pt plus 1pt   \jot = 1.5ex}
\catcode`\@=11
\def\marginnote#1{}

\newcount\hour
\newcount\minute
\newtoks\amorpm
\hour=\time\divide\hour by60
\minute=\time{\multiply\hour by60 \global\advance\minute by-\hour}
\edef\standardtime{{\ifnum\hour<12 \global\amorpm={am}%
        \else\global\amorpm={pm}\advance\hour by-12 \fi
        \ifnum\hour=0 \hour=12 \fi
        \number\hour:\ifnum\minute<10 0\fi\number\minute\the\amorpm}}
\edef\militarytime{\number\hour:\ifnum\minute<10 0\fi\number\minute}

\def\draftlabel#1{{\@bsphack\if@filesw {\let\thepage\relax
   \xdef\@gtempa{\write\@auxout{\string
      \newlabel{#1}{{\@currentlabel}{\thepage}}}}}\@gtempa
   \if@nobreak \ifvmode\nobreak\fi\fi\fi\@esphack}
        \gdef\@eqnlabel{#1}}
\def\@eqnlabel{}
\def\@vacuum{}

\def\draftmarginnote#1{\marginpar{\raggedright\scriptsize\tt#1}}

\def\draftlabel#1{{\@bsphack\if@filesw {\let\thepage\relax
   \xdef\@gtempa{\write\@auxout{\string
      \newlabel{#1}{{\@currentlabel}{\thepage}}}}}\@gtempa
   \if@nobreak \ifvmode\nobreak\fi\fi\fi\@esphack}
        \gdef\@eqnlabel{#1}}
\def\@eqnlabel{}
\def\@vacuum{}
\def\draftmarginnote#1{\marginpar{\raggedright\scriptsize\tt#1}}

\def\draft{\oddsidemargin -.5truein
        \def\@oddfoot{\sl preliminary draft \hfil
        \rm\thepage\hfil\sl\today\quad\militarytime}
        \let\@evenfoot\@oddfoot \overfullrule 3pt
        \let\label=\draftlabel
        \let\marginnote=\draftmarginnote
   \def\@eqnnum{(\theequation)\rlap{\kern\marginparsep\tt\@eqnlabel}%
\global\let\@eqnlabel\@vacuum}  }

\def\numberbysection{\@addtoreset{equation}{section}
        \def\theequation{\thesection.\arabic{equation}}}

\def\underline#1{\relax\ifmmode\@@underline#1\else
        $\@@underline{\hbox{#1}}$\relax\fi}

\def\titlepage{\@restonecolfalse\if@twocolumn\@restonecoltrue\onecolumn
     \else \newpage \fi \thispagestyle{empty}\c@page\z@
        \def\thefootnote{\fnsymbol{footnote}} }

\def\endtitlepage{\if@restonecol\twocolumn \else  \fi
        \def\thefootnote{\arabic{footnote}}
        \setcounter{footnote}{0}}  %\c@footnote\z@ }
\relax

\numberbysection
\hybrid

\def\beq{\begin{equation}}
\def\eeq{\end{equation}}

\def\G{\Gamma}
\def\g{\gamma}

\def\bea{\begin{eqnarray}}
\def\eea{\end{eqnarray}}

\newtheorem{th}{Theorem}

\begin{document}

\begin{titlepage}

\title{Tri-valent graphs and solitons}

\author{I.Krichever \thanks{Columbia University, 2990 Broadway,
New York, NY 10027, USA and Landau Institute
for Theoretical Physics, Kosygina str. 2, 117940 Moscow, Russia; e-mail:
krichev@math.columbia.edu. Research supported in part by National Science
Foundation under the grant DMS-98-02577}
\and
S.P.Novikov\thanks{University of Maryland, Colledge Park and Landau Institute
for Theoretical Physics, Kosygina str. 2, 117940 Moscow, Russia; e-mail:
novikov@ipst.umd.edu. Research supported in part by the NSF grant DMS 9704613}}
\date{September 20, 1999}
\maketitle

\begin{abstract} It is shown that a real self-adoint operator of order 4
on the tri-valent tree $\G_3$ has $(L,A,B)$-triple deformations
that preserve one energy level. Laplace type discrete symmetries
of such operators are constructed.
\end{abstract}

\end{titlepage}
\newpage

Until recently, nonlinear integrable systems have been known
for lattices $Z$ and $Z^2$, only. That are $(L,A)$-pairs for Toda type
systems for $Z$ and $(L,A,B)$-triples in the case of $Z^2$).
Discrete Eiler-Darboux and Laplace type spectral symmetries for linear
second order operators $L$ also have been known only for these lattices
(see \cite{1}). Note, that tri-valent tree $\Gamma_3$ is a discrete model of
the hyperbolic geometry of Lobachevski plane, unlike $Z^2$ which is a
discrete model of the Euclidian plane.
All previous attempts to find isospectral deformation of the second order
operators $L$ on $\Gamma_3$, even in the form of the $(L,A,B)$-triple,
$\dot L=LA-BL$, which preserve only one spectral level $L\Psi=0$,
failed (see \cite{2,3,4}).

Let $(L\Psi)_P=\sum_Qb_{P,Q}\Psi_Q$ be a linear operator on a graph.
Then the maximum diameter $max_P \  d(Q_1,Q_2)$, where $b_{P,Q_1}\neq 0,
b_{P,Q_2}\neq 0$, or $b_{Q_1,Q_2}\neq 0$, is called the order of
the equation $L\Psi=0$. The metric on the graph is defined by the condition
that the length of an edge equals 1. Here $\Psi_P$ is a function
on vertices $P$.

Let us consider a graph such that each edge has two vertices and
at each vertex three edges come together.
\begin{th} Let $L$ be a generic real self-adjoint operator of
order 4 on the tree $\G_3$. Then there exist isospectral
deformations of one energy level  $L\Psi=0$ of the form $(L,A,B)$-triple:
$$\dot L=LA-BL $$
where
$$
(L\Psi)_P=\sum b_{PP''}\Psi_{P''}+b_{PP'}\Psi_{P'}+w_P\Psi_P,
$$
$P,P',P''$ are vertices, $d(P,P'')=2,\ d(P,P')=1$, and
we assume that $b_{P,P''}>0.$
Here  $B=-A^t$, $(A\Psi)_P=\sum c_{PP'}\Psi_{P'}.$
\end{th}
The coefficients $c_{P,P'}$ for the nearest neighbours $P,P'$ are defined
as follows. Let us fix a vertex $P_0$ on $\G_3$, and  let
$R_i\in \g$ be edges of the shortest path $\g$ which connects
$P_0$ and $P$ and oriented from $P_0$ towards $P$. Let edges
$R_{i_1}',R_{i_2}'$ meet at the initial vertex
of the edge $R_i$, and edges
$R_{i_1}'',R_{i_2}''$ get off the second vertex of $R_i$.
Then the formula
$$
\chi(R_i)=-{\left(b_{R_{i_1}''R_i}\cdot b_{R_{i_2}''R_i}\right)\over
\left(b_{R_{i_1}'R_i}\cdot b_{R_{i_2}'R_i}\right)}.
$$
define multiplicative one-cocycle on $\G_3$.

The coefficients of the operator $A$ are equal
$$ c_R=-{1\over b_{R_1'R_2'}}\left(\prod_{R_i\in \g} \chi(R_i)\right),\
R=PP'.
$$
These formulas imply that the operator $LA+A^tL$ has order not greater
than $4$. Therefore,  the dynamical systems
$\dot L=LA+A^tL$ is well-defined. It has the form:
\begin{eqnarray}
\dot b_{PP''}&=&b_{P'P''}c_{P'P}+c_{P'P}b_{P'P}; \nonumber \\
\dot b_{PP'}&=&b_{P'P''_i}c_{P_i''P'}+c_{P^*_{\alpha}P}b_{P^*_{\alpha}P'}+
w_Pc_{PP'}+w_{P'}c_{P'P}; \nonumber \\
\dot w_P&=&2b_{PP'}c_{P'P}, \ \ i,\alpha=1,2.
\end{eqnarray}
Here $P_{\alpha}^*PP'P_i''$ are shortest paths of length $d=3$, that contain
the edge $PP'=R$.

{\bf Remark 1.} For any tri-valent graph $\G$ the coefficients
$c_{PP'}$ of the operator $A$ are defined on the abelian
cover of $\G$, defined by the the cocycle $\chi$.

\begin{th} A generic self-adjoint real fourth order operator $L$
on the tree  $\G_3$ admits one-parametric family
of factorizations of the form
$$
L=Q^tQ+u_P, {\rm where} \ \ (Q\psi)_{P}=\sum_Q d_{PQ}\psi_Q+v_P\psi_P,
$$
where
$$
b_{PP''}=d_{P'P}d_{P'P''}; b_{PP'}=d_{P'P}v_{P'}+d_{PP'}v_{P},
$$
$$w_P=v_P^2+\sum_{P'}d_{P'P}^2+u_P \ \  ({\rm let}\  d_{PQ}>0.)
$$
\end{th}
The coefficients $d_{PQ}$ are defined uniquely, the coefficient
$v_P$ is defined by one parameter, that is its evaluation at the initial
point $P_0\in \G_3$. The factorization define the Laplace type transform
$$\widetilde L =Qu_P^{-1}Q^t+1,\ \ \tilde \psi=Q\psi,
$$
where $\widetilde L\tilde \psi=0$, if $L\psi=0.$ The self-adjoint operator
$\widetilde L$ is defined uniqully up to the transformation
$$\widetilde L\to f_P^{-1}\cdot\widetilde L\cdot f_P ,\ \
\tilde \psi \to f_P^{-1}\cdot \tilde\psi.
$$
Let us take$f_P=u_P^{1/2}.$ Then we get
$\widetilde L=\widetilde Q^t \widetilde Q+u_P,$ where
$$\widetilde Q=u_P^{-1/2}Q^t u_P^{1/2}, \ \ \tilde \psi=u_P^{-1/2}Q\psi.$$
(cf. \cite{5} for $Z^2$).

{\bf Remark 2.} The factorization of $L$ depends on the existence
of a solution of the linear equation $b_{PQ}=d_{QP}v_{Q}+d_{PQ}v_{P}$.
Note, that this operator has non-trivial (one-dimensional)
kernel iff the cocycle $\chi$ is cohomological to zero on $\G$.

\end{document}